\documentclass[11pt]{article}
\pdfoutput=1
\PassOptionsToPackage{x11names,svgnames}{xcolor}
\usepackage[utf8]{inputenc}
\usepackage{cite}  
\usepackage{graphicx}
\usepackage[margin=1.25in]{geometry}
\usepackage{xcolor}
\usepackage{tcolorbox}
\usepackage{url}
\usepackage[breaklinks,
    colorlinks=true,
    urlcolor=blue,
    anchorcolor=blue,
    citecolor=blue,
    filecolor=blue,
    linkcolor=blue,
    menucolor=blue,
    linktocpage=true,
    pdfproducer=medialab,
    pdfa=true
]{hyperref}
\usepackage{xspace}
\usepackage{orcidlink}
\usepackage{lineno}
\usepackage{todonotes}
\usepackage[toc]{glossaries}
\usepackage{cleveref}
\usepackage[page,title,toc]{appendix}
\usepackage{breakurl}

\def\Title#1{\begin{center} {\LARGE #1 } \end{center}}

\newcommand\pubblock{\rightline{\begin{tabular}{l} \pubnumber\\
         \pubdate \end{tabular}}}

\newcommand\snowmass{\begin{center}\rule[-0.2in]{\hsize}{0.01in}\\\rule{\hsize}{0.01in}\\
\vskip 0.1in Submitted to the  Proceedings of the US Community Study\\ 
on the Future of Particle Physics (Snowmass 2021)\\ 
\rule{\hsize}{0.01in}\\\rule[+0.2in]{\hsize}{0.01in} \end{center}}

\newcommand{\rivet}{Rivet\xspace} 
 
\newcommand{\hepdata}{HEPData\xspace} 
 
\newcommand{\madanalysis}{MadAnalysis5\xspace} 
\newcommand{\checkmate}{CheckMATE\xspace}

\newcommand{\simpleanalysis}{SimpleAnalysis\xspace} 
\newcommand{\delphes}{\textsc{Delphes}\xspace} 
\newcommand{\zenodo}{Zenodo\xspace} 

\def\beq{\begin{equation}}
\def\eeq#1{\label{#1}\end{equation}}
\def\eeqn{\end{equation}}

\newenvironment{Eqnarray}%
   {\arraycolsep 0.14em\begin{eqnarray}}{\end{eqnarray}}
\def\beqa{\begin{Eqnarray}}
\def\eeqa#1{\label{#1}\end{Eqnarray}}
\def\eeqan{\end{Eqnarray}}

\let\bar=\overbar

\def\lsim{\mathrel{\raise.3ex\hbox{$<$\kern-.75em\lower1ex\hbox{$\sim$}}}}
\def\gsim{\mathrel{\raise.3ex\hbox{$>$\kern-.75em\lower1ex\hbox{$\sim$}}}}

\def\del{\partial}
\def\Dslash{\not{\hbox{\kern-4pt $D$}}}
\def\dslash{\not{\hbox{\kern-2pt $\del$}}}
\def\pslash{\not{\hbox{\kern-2pt $p$}}}
\def\ETmiss{\not{\hbox{\kern-4pt $E$}}_T}

\def\Dlr{\mathrel{\raise1.5ex\hbox{$\leftrightarrow$\kern-1em\lower1.5ex\hbox{$D$}}}}

\def\MSB{{\bar{M \kern -2pt S}}}
\def\msb{{\bar{\scriptsize M \kern -1pt S}}}

\def\drb{{\bar{\scriptsize D \kern -1pt R}}}

\newcommand\pubnumber{\hphantom{x}}
\newcommand\pubdate{\today}

\newacronym{HEP}{HEP}{High Energy Physics}
\newacronym{CERN}{CERN}{European Organization for Nuclear Research}
\newacronym{LHC}{LHC}{Large Hadron Collider}
\newacronym{HL-LHC}{HL-LHC}{High-Luminosity \gls{LHC}}
\newacronym{EIC}{EIC}{Electron-Ion Collider}
\newacronym{RHIC}{RHIC}{Relativistic Heavy Ion Collider}
\newglossaryentry{ATLAS}{name={ATLAS}, description={A general-purpose detector at the \gls{LHC}}}
\newglossaryentry{CMS}{name={CMS}, description={A general-purpose detector at the \gls{LHC}}}
\newglossaryentry{CMSSW}{name={CMSSW}, description={Offline software for the \gls{CMS} collaboration}}
\newglossaryentry{ALICE}{name={ALICE}, description={A detector specialised in heavy-ion physics at the \gls{LHC}}}
\newglossaryentry{LHCb}{name={LHCb}, description={A detector specialised in $b$-physics at the \gls{LHC}}}
\newglossaryentry{STAR}{name={STAR}, description={One of four experiments at \gls{RHIC} studying quark-gluon plasma}}
\newglossaryentry{PHENIX}{name={PHENIX}, description={Pioneering High Energy Nuclear Interaction eXperiment, was a detector at \gls{RHIC} designed to investigate high energy collisions of heavy ions and protons}}
\newglossaryentry{sPHENIX}{name={sPHENIX}, description={A detector at \gls{RHIC} studying quark-gluon plasma by combining \gls{PHENIX} and \gls{STAR}}}
\newglossaryentry{LEP}{name={LEP}, description={Large Electron-Positron Collider}}
\newglossaryentry{DESY}{name={DESY}, description={German Electron Synchrotron}}
\newglossaryentry{HERA}{name={HERA}, description={Hadron-Electron Ring Accelerator at \gls{DESY}}}
\newglossaryentry{SLAC}{name={SLAC}, description={SLAC National Accelerator Laboratory}}
\newglossaryentry{BABAR}{name={BABAR}, description={A particle physics experiment at \gls{SLAC}}}
\newglossaryentry{LZ}{name={LZ}, description={LUX-ZEPLIN dark matter experiment}}
\newglossaryentry{DUNE}{name={DUNE}, description={Deep Underground Neutrino Experiment}}
\newacronym{HSF}{HSF}{\gls{HEP} Software Foundation}
\newacronym{IRIS-HEP}{IRIS-HEP}{Institute for Research and Innovation in Software for High Energy Physics}
\newglossaryentry{data}{name={data}, description={Data recorded from experiments that have been passed through the experiment's event reconstruction}}
\newglossaryentry{derived data}{name={derived data}, description={\Gls{data} that have passed through a size reduction step that prunes information, but which might also add additional calculated quantities to the data files.}}
\newglossaryentry{data product}{name={data product}, plural={data products}, description={All files containing selections of derived data and synthesised information from the various stages of an analysis. These might include summary plots and tables, histograms of kinematic distributions, fiducial cross sections, cross section limits, simplified model results, correlation information, analysis statistical workspace binaries or full statistical models, etc}}
\newglossaryentry{data preservation}{name={data preservation}, description={The procedures, practices, and standards of ensuring the long-term (i.e., decades beyond the end of an experiment) preservation, accessibility, and usability of data and derived data from experiments}}
\newglossaryentry{analysis preservation}{name={analysis preservation}, description={The procedures, practices, and standards of ensuring the long-term preservation, accessibility, and usability of information necessary to repeat an experimental analysis (starting from its associated preserved data) and generate all associated \glspl{data product}}}
\newglossaryentry{reinterpretation}{name={reinterpretation}, plural={reinterpretations}, description={Any type of new or updated interpretation of an experimental analysis or result, including the combination in, e.g., global fits or global averages.}}
\newglossaryentry{recasting}{name={recasting}, description={Reproducing the analysis logic in a simulation, considering a different physical process with a different phase space distribution, which might have different efficiencies and acceptances than the originally hypothesised model}}
\newacronym{FAIR}{FAIR}{Findable, Accessible, Interoperable, and Reusable}
\newacronym{BSM}{BSM}{Beyond-the-Standard Model}
\newacronym{QCD}{QCD}{Quantum Chromodynamics}
\newacronym{EFT}{EFT}{Effective Field Theory}
\newacronym{ONNX}{ONNX}{Open Neural Network Exchange}
\newacronym{ML}{ML}{machine learning}
\newacronym{DSL}{DSL}{domain specific language}
\newacronym{ADL}{ADL}{analysis description language}
\newacronym{BDT}{BDT}{boosted decision tree}
\makeglossaries

\AtBeginEnvironment{appendices}{\crefalias{section}{appendix}}

\begin{document}

\pubblock
\snowmass{}

\Title{\textbf{Data and Analysis Preservation,\\ Recasting, and Reinterpretation}}

\begin{center}{\large
TF07 (Collider Phenomenology in the Theory Frontier)\\
COMPF7 (Reinterpretation and long-term preservation of data and code)}
\end{center}

\begin{center}
Stephen~Bailey\,\orcidlink{0000-0003-4162-6619}\textsuperscript{1}, 
Christian~Bierlich\,\orcidlink{0000-0002-3978-6085}\textsuperscript{2},
Andy~Buckley\,\orcidlink{0000-0001-8355-9237}\textsuperscript{3},
Jon~Butterworth\,\orcidlink{0000-0002-5905-5394}\textsuperscript{4},
Kyle~Cranmer\,\orcidlink{0000-0002-5769-7094}\textsuperscript{5},
Matthew~Feickert\,\orcidlink{0000-0003-4124-7862}\textsuperscript{6*},
Lukas~Heinrich\,\orcidlink{0000-0002-4048-7584}\textsuperscript{7},
Axel~Huebl\,\orcidlink{0000-0003-1943-7141}\textsuperscript{1},  
Sabine~Kraml\,\orcidlink{0000-0002-2613-7000}\textsuperscript{8$\ddagger$},
Anders~Kvellestad\,\orcidlink{0000-0002-5267-7705}\textsuperscript{9},
Clemens~Lange\,\orcidlink{0000-0002-3632-3157}\textsuperscript{10},
Andre~Lessa\,\orcidlink{0000-0002-5251-7891}\textsuperscript{11},
Kati~Lassila-Perini\,\orcidlink{0000-0002-5502-1795}\textsuperscript{12},
Christine~Nattrass\,\orcidlink{0000-0002-8768-6468}\textsuperscript{13},
Mark~S.~Neubauer\,\orcidlink{0000-0001-8434-9274}\textsuperscript{6},  
Sezen~Sekmen\,\orcidlink{0000-0003-1726-5681}\textsuperscript{14},
Giordon~Stark\,\orcidlink{0000-0001-6616-3433}\textsuperscript{15},
Graeme~Watt\,\orcidlink{0000-0003-0775-6604}\textsuperscript{16}
\end{center}

\begin{center}
\textbf{1}~Lawrence Berkeley National Laboratory, USA
\textbf{2}~Lund University, Lund, Sweden
\textbf{3}~University of Glasgow, UK
\textbf{4}~University College London, UK
\textbf{5}~New York University, USA
\textbf{6}~University of Illinois at Urbana-Champaign, USA
\textbf{7}~Technische Universität München, Germany
\textbf{8}~Univ. Grenoble Alpes, CNRS, Grenoble INP, LPSC-IN2P3, Grenoble, France
\textbf{9}~University of Oslo, Norway
\textbf{10}~Paul Scherrer Institute, Villigen, Switzerland
\textbf{11}~Universidade Federal do ABC, Brazil
\textbf{12}~Helsinki Institute of Physics, Finland
\textbf{13}~University of Tennessee, Knoxville, USA
\textbf{14}~Kyungpook National University, Korea
\textbf{15}~SCIPP, UC Santa Cruz, CA, USA
\textbf{16}~IPPP, Durham University, UK
\end{center}

\begin{center}
Corresponding authors:\\
*~matthew.feickert@cern.ch, $\ddagger$~sabine.kraml@lpsc.in2p3.fr
\end{center}

\begin{abstract}
\noindent We make the case for the systematic, reliable preservation of event-wise data, derived data products, and executable analysis code. This preservation enables the analyses' long-term future reuse, in order to maximise the scientific impact of publicly funded particle-physics experiments.
We cover the needs of both the experimental and theoretical particle physics communities, and outline the goals and benefits that are uniquely enabled by analysis recasting and reinterpretation. 
We also discuss technical challenges and infrastructure needs, as well as sociological challenges and changes, and give summary recommendations to the particle-physics community.
\end{abstract}

\clearpage
\tableofcontents
\clearpage

\hrule
\paragraph{Executive summary.}
To achieve their full scientific impact, HEP experiments need to integrate extensive data and analysis preservation efforts into their publication processes, alongside the communication of results in reusable form and preservation of data products, and making event-level data publicly available.
Without this, the influence of the hundreds of published analyses from the LHC, HL-LHC, EIC, and other future experiments will be limited mainly to the physics ideas in vogue at the time the collaboration collected their data.
The public investment in experimental programs underscores the importance of going beyond the original paper publication and ensuring that analyses continue providing scientific value in perpetuity.\\
\hrule

\section{Introduction}

The scientific value of the output from analyses performed at experiments in \gls{HEP} is immense.
To maximise the scientific return of the data obtained and the analyses performed at these unique machines, it is imperative that there are strategic community plans in practice for the reuse and reinterpretation of the analyses and all of their associated data products.
Enabling reuse and the ability to explore not-yet-thought-of-theories in the long-term future (i.e., on the 10 to 50-year time scale) needs to become a community priority.
The benefits extend beyond improving and extending searches for new physics, and also allow models, including new precision Standard Model calculations, to be tested in new physics regimes.
This will require reuse and preservation to become parts of the planning process for future analyses, and additionally be incorporated into the operations and facilities planning for future experiments, as well as for the phenomenology community, who will need to be major shareholders in this work.

These changes, and the technical infrastructure necessary for long-term storage and improved preservation standards, are not inconsequential or easy~\cite{LHCReinterpretationForum:2020xtr,Cranmer:2021urp}.
They will require development of standards, software, cyberinfrastructure, stewardship roles, and additional support from funding agencies.
Changes at fundamental levels to the community's approach to analysis will necessarily require additional sociological changes.
The particle physics community is not homogeneous in its experiences and practices with preservation and reinterpretation, so it is to be expected that different subfields will have different adoption times of recommendations and experience different levels of sociological challenges.
There are steps that can be taken to smooth this process in the field.
Current procedures in national laboratories and universities for publishing data and software --- e.g., by asserting copyright --- are non-uniform and often require significant overhead of individual scientists.
Reducing such administrative entry burdens to contribute to repositories by using fast and streamlined procedures,
could significantly increase contributions to data and software repositories.

In this paper, we distinguish between preserving collision and simulated event data used as input to physics analyses, described in~\Cref{data-preservation}, and preserving workflows and data products connected to specific analyses, covered in~\Cref{sec:analysis-preservation,data-products}, respectively. \Cref{reinterpret} addresses the usage of the available material for reinterpretation in terms of new theoretical ideas and global analyses.
To help with clarity of discussion, we summarise key terms used in the paper in~\Cref{fig:glossary}.
This paper focuses primarily on the \gls{LHC}, but its recommendations apply to particle physics experiments in general, and beyond physics analyses.

\begin{figure}[!ht]
\begin{tcolorbox}
\begin{center}
{\large \textbf{Glossary of terms}}
\end{center}
\begin{description}
    \item[1.1:] \textbf{\Gls{data}}: In the context of this paper, this refers explicitly to data recorded from experiments that have been passed through the experiment's event reconstruction.
    \item[1.2:] \textbf{\Gls{derived data}}: Data that have passed through a size reduction step that prunes information, but which might also add additional calculated quantities to the data files.
    \item[1.3:] \textbf{\Glspl{data product}}: All files containing selections of derived data and synthesised information from the various stages of an analysis.
These might include summary plots and tables, histograms of kinematic distributions, fiducial cross sections, cross section limits, simplified model results, correlation information, analysis statistical workspace binaries or full statistical models, etc.~\cite{Cranmer:2021urp}
    \item[1.4:] \textbf{\Gls{data preservation}}: The procedures, practices, and standards of ensuring the long-term (i.e., decades beyond the end of an experiment) preservation, accessibility, and usability of data and derived data from experiments.
    \item[1.5:] \textbf{\Gls{analysis preservation}}: The procedures, practices, and standards of ensuring the long-term preservation, accessibility, and usability of information necessary to repeat an experimental analysis (starting from its associated preserved data) and generate all associated \glspl{data product}.
    As discussed in~\Cref{sec:analysis-preservation}, there are multiple levels of analysis preservation fidelity with distinct advantages and trade offs.
    \item[1.6:] \textbf{\Gls{reinterpretation}}: Any type of new, alternative or updated interpretation of an experimental analysis or result, including the combination in, e.g., global fits or global averages. 
    \item[1.7:] \textbf{\Gls{recasting}}: Reproducing the analysis logic in a simulation, considering a different physical process with a different phase space distribution, which might have different efficiencies and acceptances than the originally hypothesised model.
\end{description}
\end{tcolorbox}
\caption{Definitions of a few key terms used in this paper.}
\label{fig:glossary}
\end{figure}

\section{Data preservation (open data)}
\label{data-preservation}

Following the positive experience and feedback from the open releases of recorded and simulated event-level datasets by the \gls{CMS} experiment since 2014, \gls{CERN} formulated an open data policy for the \gls{LHC} experiments~\cite{cern-data-policy} in 2020. As the host laboratory, \gls{CERN} maintains and develops the infrastructure needed for the portal, provides storage resources, and takes the custodial responsibility of released open data in the long term. All \gls{LHC} experiments are committed to releasing research-quality data through the \gls{CERN} Open Data portal~\cite{CODP}. The amount of data and release timeline varies from one experiment to another~\cite{cern-data-policy,cern-open-data-privacy-policy,cms-open-data-policy,atlas-open-data-policy,lhcb-open-data-policy,alice-open-data-policy}. 

\gls{CMS} is the only experiment having released research-quality data at the time of writing. The \gls{CMS} Open Data releases contain complete reprocessing of collision data from each data-taking period and the simulated data corresponding to these data. They are made available in the format and with the exact data quality requirements used by analyses of the \gls{CMS} collaboration. The volume of data, both actual and Monte Carlo simulation, amounts currently to 2.8~PB. \gls{CMS} also provides a compatible version of the \gls{CMSSW} and additional information necessary to perform a research-level physics analysis on the public data. The documentation comes with example code and specific guide pages to instruct new users.

Data released through the \gls{CERN} Open Data portal satisfy \gls{FAIR} principles for scientific data management~\cite{FAIR-paper} for data and metadata to a large extent. Due to the complexity of experimental particle physics data, the \gls{FAIR} principles alone do not guarantee the reusability of these data, and additional effort is needed to pass on the knowledge needed to use and interpret them correctly. In large experiments, collaboration members benefit from the existing knowledge infrastructure (meetings, discussion forums, mentoring, specific groups responsible for providing data assets needed for physics analysis, and more) that is unavailable to external users or long-term. 

Early testing is of utmost importance to ensure that the open data are usable for physics research and that all necessary additional information is captured, stored, and provided. Therefore, \gls{CMS} emphasises releasing data relatively early after the data-taking. External users' feedback and questions indicate the eventual missing data assets and lack of information. The release can then be complemented while the expertise on these data is still present in the collaboration. To encourage the use of open data and to get direct feedback, the \gls{CMS} Open Data group is organizing regular workshops with hands-on tutorials. It is essential to reach out to a diverse user community, from newcomers to seniors with different computing backgrounds and physics skills. While the \gls{CMS} Open Data group is looking forward to organizing the next workshop in presence, the value of remote, virtual events with many participants who would not have been able to travel is acknowledged.

Setting up example analysis workflows is one of the priorities for making open data reusable. They should demonstrate all necessary steps from data access and selection to any eventual corrections and address the particularities of experimental data, such as estimating efficiencies and uncertainties. Activities for preserving analysis workflows internally in the collaboration will facilitate building such workflows. In use with open data, the software container technology is suitable for automating workflows. The experience gained with them can be used for current analyses within the collaboration.

There have been successful data-preservation efforts in the \gls{LEP} and \gls{HERA} experiments. However, no public access to these data exists in terms defined in the \gls{CERN} data policy for the \gls{LHC} experiments. Access to preserved data is made possible through different modalities, such as joining or working with collaboration members. The \gls{BABAR} collaboration has decided to make its data available for future analyses through the \gls{CERN} Open Data portal. Lack of person-power is often a bottleneck, and while valid for experiments in the data-taking phase, it is even more evident for experiments past data-taking. Open access initiatives have to start at an early stage; see also the reports at the recent DPHEP workshop~\cite{DPHEPws:2021}.

Building a community of users for these data is paramount to making open data initiatives successful. The theory community is a critical player in this, and promoting open data despite the substantial work required in their analysis is a strong demonstration of their value and an invaluable validation effort for their usability for future generations.

\begin{figure}[!ht]
\begin{tcolorbox}
\begin{center}
{\large \textbf{Data Preservation Recommendations}}
\end{center}
\begin{description}
   \item[2.1:] Agree on \gls{data preservation} and public event-data releases as a means to maximise scientific outcomes, and allocate resources and responsibilities to achieve this goal in the experiments' organization.    \item[2.2:] Give long-term custodial responsibility of public \gls{data} to the host laboratory or an organization that persists beyond the experiment's lifetime and uses common distribution platforms with other experiments.
   \item[2.3:] Incorporate preparing \gls{data} for public releases and invest in preserving the knowledge needed for their use in the data processing and analysis operations and facilities.
   \item[2.4:] Encourage and promote the use of open data to explore and improve usability and to ensure that all necessary information for research-level use is available.
\end{description}
\end{tcolorbox}
\caption{Recommendations for data preservation.}
\label{fig:recs_opendata}
\end{figure}

\section{Analysis preservation}
\label{sec:analysis-preservation}

Preservation of analysis logic and workflows in any specific form enables the reuse of the original analysis process and associated \glspl{data product}. Such reuse may, for instance, be the \gls{reinterpretation} in terms of physics models not considered in the original analysis by the experimental collaboration; this may range from \gls{BSM} theories, to new ideas and implementations of non-perturbative \gls{QCD}, and everything in between.

As such, an experiment needs to integrate analysis preservation into its publication processes alongside open data and data-product preservation to achieve its full scientific impact. Without this, the influence of the hundreds of published analyses from the \gls{LHC}, \gls{HL-LHC}, \gls{EIC}, and other modern collider experiments will be limited mainly to the physics ideas in vogue at the time the collaboration collected collider data. The public investment in experimental programs underscores the importance of going beyond the first publication and ensuring that analyses continue providing scientific value in perpetuity.

Different levels of event-wise analysis preservation are possible.
They already exist, from full preservation of the experimental analysis workflow, to ``fast'' or ``lightweight'' emulations of the entire analysis that emphasise speed (and often greater public availability) to the expense of some precision.
Both the accurate-but-expensive and fast-but-approximate approaches (and points in between) have value for physical applications. Given an ample model space to explore, combining many preserved event analyses to obtain a more statistically significant result, the emphasis lies more on speed and efficiency than total accuracy, so a rough measure of which parameter regions are viable.
However, the emphasis naturally shifts to precision with a smaller range of models to explore, perhaps after such fast-interpretation triage.
The expense of re-running full experimental software stacks becomes justifiable and tractable.

In the following sections, we review first the full and then the lightweight preservation paradigms, and finally the current status and issues of both, in both the high-energy $pp$ and heavy-ion programs.

\subsection{Full preservation}\label{section:full-detail-preservation}

The most faithful approach to analysis preservation is to store, in a reproducible form, the exact software chain used to perform the analysis in the experiment. As each experiment has its internal data formats and software frameworks, injecting new models for analysis typically requires running the full post-generation software stack from detector simulation and reconstruction to the physics analysis code.

The obvious consequence is that simulation and reconstruction are computationally expensive, so is event reinterpretation via such \glspl{analysis preservation}. A secondary implication is for intellectual property: the ability to perform detailed simulation of an experiment's detector response is strategically sensitive information and not made explicitly usable outside the collaborations. If the code is hosted on publicly accessible platforms, this does not necessarily allow public use. Without explicit support and computing infrastructure, the complexity and resource requirements of these frameworks makes their use tractable only within the experimental collaborations.

Preserving full analysis chains is complicated by the diversity of analysis software frameworks, even within a given experiment.
Versioned central production systems within one framework typically perform event generation, simulation, reconstruction, and data reduction (``derivation'').
By contrast, most physics analyses within an experiment start from various tiers of derived data, and the lack of constraints has led to a proliferation of alternative analysis frameworks.
These often lack extensive documentation, may be version controlled in different locations, and without coordination and standardization, naturally, evolve as many incompatible technical interfaces as there are packages.

Therefore, for reproducibility, experimental collaborations and host institutions must arrange to store analysis data in well-defined repositories and structures, with a plan for long-term archiving.
The host institutions should monitor the correctness and continuing validity of the software by using continuous integration testing on a range of suitable \gls{data} for each analysis.
As a part of these tests, container images (e.g.~Docker~\cite{docker}, Podman~\cite{podman}, or Singularity~\cite{singularity}) should be built and preserved on accessible container registries to capture the software environment needed for data processing.

\subsection{Lightweight preservation}
\label{lightweight}

The approach described in~\Cref{section:full-detail-preservation} represents a faithful version of preservation, with implications for the original analysis software chain and typically high computing power requirements on any large-scale reuse of the preserved software stack. Even for organizations with the computational resources of active \gls{HEP} experiments --- where at present all such preservation has been performed --- physicists cannot use this form of preservation to explore large model-parameter spaces, e.g.~$\gg 2$ dimensions, via adaptive samplers.

These limitations motivate a more lightweight form of preservation, with more modest computing demands. It is simplest to make this alternative form independent of the original experiment software chain, with the desirable side-effect that it is also simpler to release publicly. The cost of being lightweight is that the approximations involved reduce the precision of the preservation; this makes lightweight preservation suitable for identifying model regions of interest rather than for making definitive statements of discovery. Public availability is essential for theorists' studies outside the experimental collaborations, such as testing new theoretical ideas against the data from several analyses and experiments in a global approach.

Several lightweight frameworks have arisen, ensuring not only preservation and reproducibility of experimental analyses but enabling \gls{reinterpretation} studies for the whole \gls{HEP} community~\cite{LHCReinterpretationForum:2020xtr}. The \rivet tool~\cite{Bierlich:2019rhm} is the clear choice for (differential) measurements, particularly where detector effects have been unfolded to a fiducial phase-space by the original experiment (and the statistical consequences fully reported, cf.~\Cref{data-products}). The \gls{LHC} experiments have established \rivet-based measurement-preservation programs, and similar efforts are gaining traction at the \gls{RHIC}, e.g.~\gls{STAR} and \gls{PHENIX} programmes based around \rivet's heavy-ion features~\cite{Bierlich:2020wms}, and are being planned at the \gls{EIC}. Such initiatives need to be built into the data publication and exploitation plans of ongoing and future experiments, to maximise the scientific impact of the analyses.

Analysis preservation at the reconstruction-level is more ambiguous, as some form of detector response needs to be encoded, and this will necessarily be less accurate than the original detector simulation+reconstruction code.
\rivet incorporates a transfer-function approach for experiment-provided efficiencies and kinematic distortions to be applied to generator-level events but currently contains few reconstruction-level (usually \gls{BSM} searches) analyses.

In parallel, driven by the need of open access to \gls{BSM} recasting, the theory community has been developing various simulation-based reinterpretation frameworks for reconstruction-level analyses, in particular \checkmate~\cite{Drees:2013wra,Dercks:2016npn}, \madanalysis~\cite{Dumont:2014tja,Conte:2018vmg} and GAMBIT's ColliderBit~\cite{GAMBIT:2017qxg}. They either also use a transfer-function approach, or rely on \delphes~\cite{deFavereau:2013fsa} for the emulation of detector effects; in some cases also generator-level events are used. A detailed overview of approaches and public frameworks is given in \cite{LHCReinterpretationForum:2020xtr}.
Experiments can also publish analyses in lightweight, executable form such as \gls{ATLAS}' semi-public \simpleanalysis system~\cite{atlas:simpleanalysis}
(describing in particular, the analysis logic but not necessarily reconstruction efficiencies, and more.)

As in the cases of full analysis preservation and \rivet-based lightweight measurement preservation, active policy and support efforts will be required in experiments as part of the publication process to ensure the required level of preservation coverage. An example of good practice in this respect is the \gls{CMS} jets+missing energy analysis \cite{CMS:2021far}, which provided a \madanalysis recast code \cite{Albert:2774586,DVN/IRF7ZL_2021} together with the paper, and in \gls{ATLAS} and \gls{CMS}' established integration of \rivet-analysis release into the publication process of suitable data-analyses.

All current tools support standard MC-generator tools such as the HepMC event record. However, essential features such as propagation of systematic uncertainties via weight vectors are not wholly or consistently implemented, and MC generators also have not yet standardised weight-coding conventions. Placing emphasis again on standard interfaces will help speed the convergence of the reconstruction-level \gls{BSM} search tools in particular.

While public, lightweight preservation of cut-and-count analyses is increasingly becoming standard, the preservation of analyses that employ \gls{ML} is still in its infancy.
Indeed there are very few examples of analyses where communicating machine-learned models has been tried. In particular, the \gls{CMS} all-hadronic search for supersymmetry~\cite{CMS:2017qxu} published a simplified version of their top-quark tagger~\cite{cms:toptagger}, which is based on a Random Forest decision tree, and the \gls{ATLAS} 0-lepton gluino/squark search~\cite{ATLAS:2020syg,Uno:2763449} published the \gls{BDT} weights of their event selection as \texttt{XML} files on \hepdata~\cite{hepdata.95664.v2/r8}.

Standards exist for a degree of \gls{ML} interoperability, including the exchange of neural networks and \glspl{BDT}, e.g.~the
\gls{ONNX}~\cite{onnx} format or direct preservation of decision trees as framework-independent code, but their long-term stability is unclear; the inclusion of implementation-specific behaviours mean that for long-term preservation either exact versions of frameworks (often Python-specific) need to be re-used, or functionality limited to a minimal subset. Trained networks have bee published in \gls{ONNX} format by the \gls{ATLAS} search for $R$-parity-violating supersymmetry~\cite{ATLAS:2021fbt,hepdata.104860.v1/r3}; this is also included in the \gls{ATLAS} \simpleanalysis framework~\cite{atlas:simpleanalysis}.
However, detailed documentation of, e.g.~the input variables is missing, and it is unclear how to verify that physics objects from any given fast-simulation package will produce the intended \gls{ML} responses within acceptable uncertainties.

Further development along these lines is highly needed. Successful sharing of a \gls{ML} model requires not only sharing the model itself (including architectures, weights, and complete specification of software dependencies) but also the detailed specification of the input data. See also the corresponding discussion in Ref.~\cite{snowmass:MLevts} in this context.

Finally, a better meta-data stewardship for preserved analysis codes is in order, to make them searchable and findable (see also \Cref{reinterpret} in this context).

\subsection{Analysis description languages}

The full and lightweight preservation frameworks described above constitute a crucial step in analysis preservation, with the significant benefit of being executable.  
However, an equally important step is to preserve analysis logic in a more accessible and easily communicable, yet unambiguous format.%
\footnote{As should be the case in the paper publication, but rarely is given to the level of detail required for full reproduction.}
The thousands of analyses designed by experimental collaborations and phenomenologists have accumulated a tremendous source of physics content. This diverse content can inspire and inform new analysis ideas or train the new generation of particle physicists.
Therefore, the third facet of \gls{analysis preservation} is to encode an abstract description of the analysis logic in a \emph{human- and machine-readable} declarative language. 

Physics analyses typically consist of multiple, separately executed steps (e.g.~event, selection and signal extraction). It is also important to capture the workflows for running and combining them over multiple event samples. There is currently no \emph{de~facto} standard for this workflow language across the experimental community: for example, Common Workflow Language~\cite{CWL}, Snakemake~\cite{SnakeMake}, Yadage~\cite{Cranmer:2017frf,yadage_code}, and Argo Workflows~\cite{argo} are in use by different collaborations. Standardization and evolution around a smaller (or unique) subset of such languages would improve interoperability and knowledge transfer. In making such a choice, we stress the importance of a \textsl{declarative} rather than \textsl{imperative} model~\cite{10.3389/fdata.2021.661501}, as the former enables researchers to concentrate on the physics task with minimal need to consider technical details such as scalable job orchestration. Developers can significantly reduce the complexity of workflow descriptions by establishing standard interfaces for tools implementing processing steps so that they can be more easily composed.

As with declarative workflow description formats, a recent trend is to decouple analysis logic from the software frameworks used to execute it. Analysis frameworks internally used in \gls{ATLAS} and \gls{CMS}
are moving towards accumulating all information related to object and event processing in a single location, e.g.~in a configuration file, at least for part of the processing chain. An approach that takes this one step further is to use \glspl{DSL} dedicated to expressing the physics content of \gls{HEP} analyses. \glspl{DSL} can either rely on the syntax of an existing computer language (embedded \gls{DSL}) or have a custom syntax (external \gls{DSL}) more tailored to the semantics of the \gls{HEP}-analysis context. Here also, a declarative approach has many benefits due to abstracting practical details into framework implementations. Several \gls{DSL} approaches are studied, and various working prototypes exist.

On the embedded \gls{DSL} side, the \texttt{F.A.S.T.} framework~\cite{FAST}, used in \gls{CMS}, \gls{LZ}, and \gls{DUNE} analyses, incorporates a \texttt{YAML} \gls{DSL}.
More recent Python based developments include \texttt{NAIL}~\cite{NAIL}, used in the \gls{CMS} analysis that observed the first evidence for Higgs to two muon decays~\cite{CMS:2020xwi}, and bamboo~\cite{David:2021ohq} which has been used in studies for Snowmass.
Additionally, FuncADL~\cite{Proffitt:2021wfh} constructs hierarchical data queries using SQL-like concepts in Python and has been used to explore applications of functional programming.

On the external \gls{DSL} side, the most advanced example is the so-called \gls{ADL}%
~\cite{adlweb, Unel:2021edl}, see also \cite{LHCReinterpretationForum:2020xtr}. 
It organizes the analysis description in multiple blocks separating object, variable and event selection definitions, and has keywords specifying analysis concepts and operations. 
\gls{ADL} can be executed by any framework capable of parsing its syntax, notably the runtime interpreter CutLang~\cite{Sekmen:2018ehb}. \gls{ADL} together with CutLang thus provide a functioning example of lightweight preservation in the sense of \Cref{lightweight}. 
The \gls{LHC} analyses implemented by \gls{ADL} are currently preserved in a GitHub repository~\cite{adllhcanl}.  \gls{ADL} principles and prospects are discussed further in a dedicated Snowmass White Paper contribution~\cite{snowmass:ADL}.

One caveat to note is that \gls{HEP} analyses are not always fully domain-specific and often need to execute custom logic in addition to predefined behaviors.
This could be remedied either by extending the \gls{DSL} to incorporate new known behaviors via an ``escape'' mechanism to inject general-purpose code (compromising its framework independence), or combining them with pre-and post-processing code to cover gaps in the \gls{DSL} capabilities.
\gls{ADL}  follows the former approach, and relies on an external database of injectable functions for each target framework.

The applications of \gls{ADL} and \gls{ADL}-type preservation are numerous. It may be used for analysis design, be directly included in publications, serve for pedagogical purposes and knowledge transfer, 
act as a configuration file that transpiles to code in lightweight analysis preservation frameworks, or
as a source to transpile between full and lightweight preservation frameworks.

\begin{figure}[ht!]
\begin{tcolorbox}
\begin{center}
{\large \textbf{Analysis Preservation Recommendations}}
\end{center}
\begin{description}
   \item[3.1:] Ensure use of interoperable systems to maximise the preservability and reusablility of experiment simulation and analysis software chains. This includes the use of version control, archival systems, containerisation, common software interfaces and data formats, and commitments from experimental collaborations and their host laboratories to maintain documentation and provide long-term support.
   
   \item[3.2:] Ensure that all operational and in-preparation experiments have a planned and resourced programme for capture and long-term reproduction of their complete computational processing chain, including validation regression-tests.
   
   \item[3.3:] Ensure that release of \gls{analysis preservation} logic via public frameworks for the community to use is integrated with experiment publication and data-release processes, to maximise analysis impact. This also includes providing clear documentation and making all dependent frameworks available and documented for community consumption.

   \item[3.4:] Support continuing development and uptake of new technologies for increasingly framework-independent analysis specifications, such as via declarative domain-specific analysis description languages.
\end{description}
\end{tcolorbox}
\caption{Recommendations for analysis preservation.}
\label{fig:recs_analysispreservation}
\end{figure}

\section{Preservation of data products}
\label{data-products}

There is widespread consensus in the community that experiments should systematically provide all relevant derived data and \glspl{data product} in an open-access numerical form for future reuse~\cite{LHCReinterpretationForum:2020xtr}. 
Such \glspl{data product} are also called publication-related or ``Level 1'' data in the 
\href{https://opendata.cern.ch/docs/about}{DPHEP categories}. 
They include observed and expected event counts  (including error sources), efficiency functions,  bin-to-bin correlations, profile likelihoods, full statistical models, signal efficiencies, simplified model results, and much more, 
as discussed extensively in Refs.~\cite{LHCReinterpretationForum:2020xtr,Cranmer:2021urp}.

With regards to publication infrastructure, 
\hepdata~\cite{hepdata} is the primary open-access repository for data products from particle physics experiments, with a long history going back to the 1970s.
Funding is provided by the UK Science and Technology Facilities Council (STFC) to Durham University (UK) for staff to maintain the operation of the \href{https://www.hepdata.net/}{hepdata.net} site, provide user support, and develop the open source software (available on the \href{https://github.com/HEPData}{\hepdata GitHub organisation}) underlying the web application.

In the past, \hepdata staff at Durham University handled data preparation in a standard format and uploaded it to the repository.
However, now these tasks are delegated to the experimental collaborations.
Data submitted to \hepdata (as \texttt{YAML}) is primarily in a tabular form that can be interactively plotted in the web application and automatically converted to other standard formats (i.e. \texttt{CSV}, \texttt{JSON}, \texttt{ROOT}, \texttt{YODA}).
The interactive nature of \hepdata means that data tables must be kept sufficiently small ($\sim$MB or less) that they can render in a web browser.
In practice, tables with more than $\sim 10,000$ rows (for example, a covariance matrix for a measurement with $\sim 100$ bins) cannot efficiently render in a web browser.
However, moderately large tables or non-tabular data files can be attached to a \hepdata record as additional resources (in any format).
The original files are downloadable, but the interactive nature is lost.
\hepdata imposes an overall size limit of 50 MB on the uploaded archive file to avoid problems caused by the attempted upload. 
Data products that are not suitable for \hepdata, due to either being too large or predominantly in a non-tabular format, might be submitted to another data repository like \href{https://zenodo.org/}{Zenodo}.
Zenodo currently plugs a gap to host data (and software) that do not fit into other repositories.%
\footnote{This concerns also model files, Monte Carlo simulation, and data products from phenomenological studies.}
A new \gls{HEP}-specific instance of Zenodo could perhaps better serve the particle physics community in the future.

Reporting of results on \hepdata has become a standard procedure in the \gls{LHC} community, with \gls{ATLAS}, \gls{CMS}, \gls{ALICE}, and \gls{LHCb} all providing the results and data products from publications.
This is also increasingly the case in the heavy-ion community with the \gls{STAR}, \gls{PHENIX}, and NA61/SHINE collaborations publishing their data products to \hepdata as well in recent years.
However, as use of \hepdata is not a particle physics wide community norm yet, coverage remains incomplete\footnote{Even in \gls{ATLAS} and \gls{CMS}.} and will continue to until there are cultural shifts, as the heavy ion community has recently experienced.

Often, instead of being provided on \hepdata, digitised results/plots are available only on collaboration web pages, without appropriate documentation, versioning, or other data stewardship.
Sometimes, the linked ROOT files are wrong (not corresponding to the associated plot) or otherwise corrupted.
In some cases, the digital material is missing altogether.
Experience shows that missing or wrong resources can rarely be retrieved or corrected, as often the analysis team has disbanded with analyzers leaving the field, or the relevant files become lost. 
Part of the problem is missing time and community recognition to providing material on \hepdata.
The \href{https://indico.cern.ch/category/14155/}{RAMP} (``Reinterpretation: Auxiliary Material Presentation'') seminar series aims at providing more visibility and recognition for such efforts,  but more is needed for a 
sustainable change of culture. 

As with \gls{analysis preservation}, current heavy-ion experiments have less comprehensively established procedures for data-product preservation in \hepdata than their \gls{LHC} counterparts, but such procedures are now integrated into the publication processes for \gls{STAR} and \gls{PHENIX}, and decoupled from publication in \gls{sPHENIX}. Coverage and process for BRAHMS and Phobos is less developed. \gls{STAR} and \gls{PHENIX} have both also instituted programmes of transcription of previously published data to \hepdata, including as a form of experimental shift in 2020--22. Ensuring systematic and sufficiently detailed preservation of analysis products from all experiments in a central subject database such as \hepdata is a central component of maximising scientific impact and data re-use, and should be designed into new experiment investment and deliverables from the start. 

Regarding \hepdata itself, material beyond digitised plots (like MC run cards, input files for benchmark points, or, most importantly, statistical models) are ``additional resources'', often lumped together in compressed archives without any standard structure.
The types of data products being preserved has become much richer and diverse than flat tables.
To be able to provide the necessary infrastructure for all of these data products will require additional funding.
There is room, and a clear need, for \gls{FAIR}-ification of these precious material.

\begin{figure}[!ht]
\begin{tcolorbox}
\begin{center}
{\large \textbf{Data Product Preservation Recommendations}}
\end{center}
\begin{description}
   \item[4.1:] Make the provisioning of all \glspl{data product} associated with an experimental analysis a mandatory step for publication.
   Establishing appropriate person power, time, and community recognition is essential to that end.
   \item[4.2:] Assure appropriate resources and funding for further development of the cyberinfrastructure, such as \href{https://www.hepdata.net/}{\hepdata} and other repositories like \href{https://zenodo.org}{Zenodo}, to preserve the data products and metadata, and extend the current data structure to include more rich data products and information beyond paper plots and flat tables, e.g., statistical models, in an individually searchable and citeable form.
\end{description}
\end{tcolorbox}
\caption{Recommendations for data product preservation.}
\label{fig:recs_datapreservation}
\end{figure}

\section{Reinterpretation and recasting}
\label{reinterpret}

The physics impact of an experimental analysis can be increased well beyond its original purpose through \gls{reinterpretation}.
The various kinds of \gls{reinterpretation} include:
\begin{description}
	\item{\it updates of existing analyses} using e.g.\ more precise theoretical calculations, improved experimental calibrations, or a different probability model;
	\item{\it parametric reinterpretation} reparametrizing the likelihood through rescaling, without altering the efficiencies and acceptances that might modify the distributions --- this is the approach taken for example when reusing simplified model results from the LHC, or in in the context of Higgs signal strengths; 
    \item{\it kinematic reinterpretation} considering a different physical process with a different phase space distribution, which might have different efficiencies and acceptances --- this is what we generally refer to as \gls{recasting}; a concrete example from these proceedings is the reuse of multi-boson and top-quark measurements to constrain new scalar states in composite Higgs models in \cite{Banerjee:2022xmu}; 
    \item{\it combinations of analyses or datasets} in model surveys,%
    \footnote{Reinterpretation of several analyses within a given (usually \gls{BSM}) scenario is relevant for displaying the complementary of distinct searches as well as identifying possible gaps in coverage. Such gaps can then be used as motivation for designing new experimental searches.} 
    global fits or global averages; this includes global \gls{EFT} and \gls{BSM} analyses as well as the reuse of datasets for the determination of parton distribution functions (discussed in more detail in \cite{Cranmer:2021urp}).
\end{description}

For this to be possible, the preservation of \glspl{data product} and analyses (usually in a lightweight format) are essential, as discussed in~\Cref{sec:analysis-preservation,data-products}.
It is relevant to note here, that for many purposes of \gls{recasting},\footnote{Except, for instance, when signal-background interference effects play a role.} analyses need only be preserved to the extent that the new signal yields can be determined and the subsequent statistical analysis using them can be performed. In this case, details of e.g.\ the derivation of background estimates need not be captured as they should already be preserved in the \glspl{data product} described in Section~\ref{data-products}.

\Glspl{reinterpretation} are most often done by physicists (from both, the theory and the experiment sides) outside the experimental collaborations, but sometimes also collaboration-internally. 
In \gls{ATLAS}, \gls{analysis preservation} for reinterpretation has been more seriously pursued in recent years. A large number of analyses is preserved using Docker images and the yadage workflow language~\cite{Cranmer:2017frf,Cranmer_2015}. This has led to the first successful uses of the RECAST~\cite{Cranmer:2010hk} paradigm, in which existing analyses are reinterpreted at full fidelity within the collaboration~\cite{RECAST1, RECAST2, RECAST3}. Currently the selection of candidate reinterpretations is done mostly within the experiment. In the future, a public portal in which the wider \gls{HEP} community has access to the catalog of RECAST-able analyses and can provide input for future full-fidelity reinterpretations will be desirable. Reinterpretations like such provide additional scientific value and should be published in peer-reviewed venues in the future. As such, they are also producers of \glspl{data product} in their own right. Products such as yields of the newly studied signal, statistical model fragments (``patches''), etc., should be submitted to archives such as \hepdata or \zenodo as well.

Within \gls{CMS}, reinterpretation or recasting is largely performed within ongoing analyses or via statistically combining analyses that explore complementary final states. 
While the analysis code is not systematically archived, the \glspl{data product} for signal extraction are usually preserved so that uncertainties across different analyses can be treated consistently in a statistical combination. 
We refer to Section~4 of \cite{Cranmer:2021urp} for more discussion of collaboration practices and use cases, including inter-experiment combinations, \gls{EFT} fits, measurements in the flavour sector, etc..

Outside of the collaborations several software tools are being developed (CheckMATE~\cite{Drees:2013wra,Dercks:2016npn}, MadAnalysis5~\cite{Dumont:2014tja,Conte:2018vmg}, Contur/Rivet~\cite{Buckley:2021neu}, SModelS~\cite{Kraml:2013mwa,Alguero:2021dig}, and others as detailed in Section~III of Ref.~\cite{LHCReinterpretationForum:2020xtr})
and are publicly available for the task of reinterpretation and/or recasting. 
They typically supply a database of implemented analyses, but also allow the user to implement new ones. 
In addition, they provide functionality for making statistical statements about the results, either by implementing the statistical models supplied by the collaboration (if available) or by taking some simplifying assumptions. We note that 
maintaining these tools and implementing new analyses requires considerable person power and funding. 

The public tools provide distinct analyses coverage and sometimes different implementations of the same analysis. 
The proliferation of analyses and tools, and the lack of interoperability between the tools, can make the complete coverage of a physics case in reinterpretation studies seriously difficult. 
A unified format for analysis implementation, which could be used interchangeably by the different tools, would significantly improve the reinterpretation potential of the phenomenology community. 
Even though this proposal has been discussed in the past~\cite{lukas_heinrich_2017_6362700}, not much progress has been made in this direction. A possible interface format to make recast codes interchangeable between frameworks  might be based on \gls{ADL}~\cite{adlweb, Unel:2021edl} if parsers for the most common public frameworks (best including automatised validation) are developed.  

Meanwhile, a few steps could be taken by the community with immediate benefits. One is a centralised (meta)database where the analyses available in the specific tools and the corresponding validation material can easily be found.%
\footnote{In this regard, the  long-lived particle (LLP) \gls{recasting} community created a centralised location for stand-alone LLP \gls{analysis preservation} and validation material in form of a GitHub repository~\cite{llpRepo}.}
A searchable database covering all tools, including the major reinterpretation frameworks, and all analysis types, would clearly be of great benefit. 
In addition, whenever possible, it would be helpful to adopt a few validation guidelines to allow for a proper estimate of the recasting uncertainties introduced by each analysis implementation.
As a second step, adopting basic standards for the input and output formats would also help the user to efficiently use distinct tools.

Another vital aspect of reinterpretation is the statistical treatment of the results. In recent years, the amount of information provided by experimental collaborations has increased significantly, allowing for a more robust statistical interpretation in phenomenological studies. However, to take full advantage of these new developments, it would be desirable to coordinate and unify the statistical output format and treatment within the specific recasting tools to have a common ground for comparison. Another critical aspect to be taken into account is the possibility for a global analysis of the results. 
Such global approaches are attempted by, e.g., the GAMBIT collaboration~\cite{GAMBIT:2018gjo,Kvellestad:2019vxm} and the ``protomodelling'' project in \cite{Waltenberger:2020ygp}.
They could, in principle, also combine results from different recasting tools.
To help in this direction, the statistical analysis should be factorised as much as possible within each tool, allowing again for interoperability. In addition,  standards and guidelines should be established for presenting the results and providing the required output for the statistical interpretation.

A further motivation for facilitating collider reinterpretations in global analyses is that it enables large-scale and adaptive exploration of the complementarities between collider results and other experimental results. Such global fits have the potential of both uncovering gaps in the experimental coverage, and identify which uncovered \gls{BSM} scenarios are most plausible in light of other experiments, and thus constitute well-motivated targets for future analyses/experiments. Realising this potential, however, depends critically on computational efficiency, since a global fit faces the additional computational cost of reinterpreting all relevant non-collider results and explore a typically many-dimensional theory parameter space. As such, a focus on code efficiency, stability and parallelisability in reinterpretation tools, and on development of fast approximations for expensive computations (e.g.~computation of higher-order cross-sections), is important to enable proper utilisation of experimental results.

Last but not least, it is important to ensure reproducibility and  preservation of the results obtained.
The same platforms used for \gls{data product} preservation (see~\Cref{data-products}) can also be used by the theory community to preserve the reinterpretation results and data.
In particular, \href{https://zenodo.org/}{Zenodo} has already been used by a few groups to publish auxiliary material from phenomenological studies, see e.g.~\cite{zenodo:LHCreinterpretation,zenodo:gambit}.
On the code side, is imperative for the purpose of reproducibility that the tools used for obtaining the results are properly documented and \emph{versioned}.
This policy should be largely encouraged within the theory community.\\

\begin{figure}[!ht]
\begin{tcolorbox}
\begin{center}
{\large \textbf{Reinterpretation and Recasting Recommendations}}
\end{center}
\begin{description}
    \item[5.1:] Encourage that reinterpretability and reuse be kept in mind early on in the analysis design.
    This concerns, for instance, the choice of input parameters in \gls{ML} models, the full specification of the fiducial phase space of a measurement in terms of the final state, including any vetos applied, and generally the choice of non-overlapping regions and standard naming of shared nuisances to facilitate the combination of analyses.
    \item[5.2:] Design the format and nature of the public and internally preserved \glspl{data product}, such as statistical models, with reinterpretation use-cases in mind.
    \item[5.3:] Improve the coordination among the different public reinterpretation frameworks  with the goal of a centralised database of recast codes, common input/output formats, and a unified statistical treatment.
    \item[5.4:] Encourage the \gls{FAIR}-ification of codes and \glspl{data product} from (theory) reinterpretation studies outside the experimental collaborations at the same level of sophistication as asked for experimental analyses and results. Suitable repositories are, e.g., GitHub and Zenodo; appropriate versioning is essential.
\end{description}
\end{tcolorbox}
\caption{Recommendations for reinterpretation and recasting.}
\label{fig:recs_reinterpretation}
\end{figure}

\section{Conclusions}

The recommendations we put forth in this paper are designed to give the particle physics community actionable steps in ensuring robust preservation of data, analysis logic, and tools that will allow us to get the most scientific value possible out of the rich data and data products from the experiments.
One primary goal is to reduce the amount of overhead by our colleagues in \gls{HEP}, either in re-reproducing existing analysis logic, or having to re-derive or re-produce data products that should have already been made more accessible from the start.
For example, by looking towards harmonisation of analysis logic, experimental collaborations could begin automatising the data products necessary for reinterpreation and reusability, and reduce the burden on their colleagues.

Our recommendations in~\Cref{fig:recs_opendata,fig:recs_analysispreservation,fig:recs_datapreservation,fig:recs_reinterpretation} will ideally be enacted early on in the design and planning phases for experiments and analyses, and provisioned with appropriate resources and funding. 
If adopted rigorously, our guidelines will allow for new scientific results, including testing of new models, for decades to come through reinterpretation of published experimental analyses and results, much beyond the lifetime of the experiment.

\section*{Acknowledgments}
\addcontentsline{toc}{section}{Acknowledgments}

Andy Buckley, Jon Butterworth, and Graeme Watt are supported by the UK STFC Consolidated Grant programme.
Kyle Cranmer, Matthew Feickert, and Mark Neubauer are supported in part by the National Science Foundation under Cooperative Agreement OAC-1836650.
Lukas Heinrich is supported by the Excellence Cluster ORIGINS, which is funded by the Deutsche Forschungsgemeinschaft (DFG, German Research Foundation) under Germany’s Excellence Strategy - EXC-2094-390783311.
Axel Huebl acknowledges support by the Exascale Computing Project (17-SC-20-SC), a collaborative effort of the U.S. Department of Energy Office of Science and the National Nuclear Security Administration.
Sabine Kraml acknowledges support by the IN2P3 master project ``Th\'eorie -- BSMGA'' and the joint ANR-FWF project PRCI SLDNP grant no.~ANR-21-CE31-0023.
Andre Lessa is supported by Sao Paulo Research Foundation (FAPESP) grants no.~2018/25225-9 and 2021/01089-1.
Christine Nattrass acknowledges support from the Division of Nuclear Physics of the U.S. Department of Energy under Grant No. DE-FG02-96ER40982 and from the National Science Foundation under Grant No. OAC-1550300.
Sezen Sekmen is supported by the Basic Science Research Program through the National Research Foundation of Korea (NRF) funded by the Ministry of Education under contracts NRF-2021R1I1A3048138, NRF-2018R1A6A1A06024970 and NRF-2008-00460. 

\def\thefootnote{\fnsymbol{footnote}}
\setcounter{footnote}{0}

\addcontentsline{toc}{section}{References}
\bibliographystyle{JHEP}
\bibliography{bib/preservation,bib/reinterpretation}

\clearpage
\printglossaries

\end{document}